\documentclass[aps,pra,twocolumn]{revtex4-1}
\usepackage{graphicx}

\newcommand{\ket}[1]{\left\vert{#1}\right\rangle}
\usepackage{graphicx}
\usepackage{color}
\begin{document}
\title{Implementation of the quantum
Fourier transform on a hybrid 
qubit-qutrit NMR quantum emulator}
\author{Shruti Dogra}
\email{shrutidogra@iisermohali.ac.in}
\address{Department of Physical Sciences,
Indian
Institute of Science Education \&
Research Mohali, Sector 81 Mohali,
Manauli PO 140306 Punjab India.}
\author{Arvind}
\email{arvind@iisermohali.ac.in}
\address{Department of Physical Sciences,
Indian
Institute of Science Education \&
Research Mohali, Sector 81 Mohali,
Manauli PO 140306 Punjab India.}
\author{Kavita Dorai}
\email{kavita@iisermohali.ac.in}
\address{Department of Physical Sciences,
Indian
Institute of Science Education \&
Research Mohali, Sector 81 Mohali,
Manauli PO 140306 Punjab India.}
\begin{abstract}
The quantum Fourier transform (QFT) is a key
ingredient of several quantum algorithms and a
qudit-specific implementation of the QFT is hence
an important step toward the realization of
qudit-based quantum computers.  This work develops
a circuit decomposition of the QFT for hybrid
qudits based on generalized Hadamard and
generalized controlled-phase gates, which can be
implemented using selective rotations in NMR.  We
experimentally implement the hybrid qudit QFT on
an NMR quantum emulator, which uses four qubits to
emulate a single qutrit coupled to two qubits. 
\end{abstract}
\maketitle
\section{Introduction} 
\label{intro}
The size of a quantum register can be increased either by
increasing the number of qubits (in the standard model of a
quantum computer) or by increasing the number of accessible
logical states in each quantum element i.e by using qudits
(quantum digits)~\cite{greentree-prl-04,lanyon-naturephy-09,
brennen-pra-05,klimov-pra-03,lin-pra-09}.  For qudit
computing to become a reality, it is essential to concretize
the theoretical framework for computation in terms of a set
of quantum gates that are universal and can implement any
algorithm efficiently.  It is however not always
straightforward to generalize a qubit gate to the qudit
scenario and in some cases, more than one kind of
generalization is possible, each retaining some features of
the original qubit gate~\cite{muthu-pra-00}.  A gate library
for qudits has been proposed, containing one-qudit and
two-qudit gates such as the CINC gate, generalized SWAP and
general controlled-X (GCX)
gates~\cite{brennen-qic-2006,bullock-prl-2005,
wilmott-ijqi-12,di-pra-13,garcia-qip-13}.  Several qudit
computing proposals using different physical systems have
been designed to incorporate ``hybrid'' qudits of different
dimensions~\cite{khan-comp-06,li-pra-13,rousseaux-pra-13},
wherein multi-qudit gates are generalized to transform two
or more qudits of different dimensions.  A set of hybrid
quantum gates have been designed to act on qudits of
different dimensions, including the  hybrid SUM, SWAP,
Toffoli and Fredkin gates~\cite{daboul-jpa-03,gottesman}.
Qudits have been used as a resource in secure quantum
communication~\cite{peres-prl-00}, quantum key
distribution~\cite{groblacher-njp-06}, quantum
computing~\cite{leary-pra-06} and quantum
cloning~\cite{nagali-prl-10}.  

NMR has been a fruitful testbed for implementing
quantum information processing ideas and is now
increasingly being used to manipulate qudits.
NMR qudit implementations
include using quadrupolar nuclei oriented in a liquid
crystal~\cite{das-ijqi-2003} for
information processing, studying the
dynamics of nonclassical correlations~\cite{pinto-pra}
and finding the parity of a
permutation using a single qutrit~\cite{dogra-pla-14}
and a single ququart~\cite{pinto-qph}.
Logic gates and pseudopure states for a ququart
have been implemented using ${}^{23}$Na and
${}^{7}$Li nuclei oriented in a liquid-crystalline
matrix~\cite{khitrin-jcp,neeraj-jcp}.

The quantum Fourier transform (QFT)
needs to be specifically tailored for
qudit computers  
as it plays a key role in several quantum algorithms such as
factorization~\cite{shor-sjc-1997}, quantum phase
estimation~\cite{cleve-proc} and the hidden subgroup
problem~\cite{sim-sjc-1997}.  An efficient decomposition of
the QFT for qubits achieves an exponential speedup over the
classical fast Fourier transform~\cite{coppersmith-94},  and
has been experimentally implemented by several
groups on different
physical systems~\cite{weinstein-prl-01,dorai-ijqi-05,wang-jphysb-11,
obada-josab-13,dong-josab-13}.
Decompositions of the QFT for qudit systems 
have been worked out by several
groups~\cite{muthu-jmo-02,zobov-jetp-06,zilic-ieee-07,ye-comm-11}.

In NMR systems with ``strong'' spin-spin coupling terms in
the Hamiltonian, the identification of spins as qubits
becomes problematic. In such cases, quantum gates have been
implemented by selectively manipulating individual
transitions between pairs of energy
levels~\cite{gopinath-pra-2006,lee-apl-2006}.  Such
``effective'' qubits have been constructed by a mapping to
the non-degenerate energy eigenstates of a single qudit and
have been experimentally demonstrated by partially orienting
a ${}^{133}$Cs nuclear spin (S=7/2) in a liquid-crystalline
medium~\cite{khitrin-pra,neeraj-pra,pinto-ijqi}.  We carry
these ideas forward by carving out a system of two qubits
and a qutrit (henceforth referred to as a QQT system) out of
four qubits (henceforth referred to as a QQQQ system).
Specifically we use the four coupled qubits of
5-Fluorotryptophan to emulate a hybrid system of two qubits
and a qutrit (a $2 \otimes 2 \otimes 3$ dimensional Hilbert
space).  The qubit architecture of 5-Fluorotryptophan due to
its specific spin-spin coupling strengths and molecular
symmetries, gives rise to twenty lines in the NMR spectrum
instead of thirty two lines expected for the non-overlapping
energy levels of four qubits.  This matches the spectrum
expected for a system of two qubits coupled to a qutrit.
This system obviously cannot be used as a full-fledged QQQQ
system of four qubits due to the degeneracy of transitions.
However, by re-labeling and creating a mapping of
corresponding energy levels, we are able to show that this
QQQQ system can indeed emulate a QQT system and that the
allowed non-degenerate transitions are sufficient to
implement the required quantum gates.

In this paper we describe a decomposition of the hybrid
qudit QFT, using a set of generalized Hadamard and
generalized hybrid controlled-rotation gates.  
The scheme is general and can be implemented on
any physical hardware for a hybrid qudit quantum computer.
We implement the hybrid QFT on a QQT system (using the four
qubits of 5-Fluorotryptophan as a QQT emulator).  
The requisite hybrid
qudit gates can be realized in NMR using selective
rotations.  
Partial
state tomography of the final state was achieved by a set of
19 specially designed experiments to reconstruct desired
portions of the Hilbert space. The results of the tomography
demonstrate the success of the mapping of the QQT system
onto the QQQQ system as well as the implementation of the
hybrid qubit-qutrit QFT on the system.

The material in this paper is organized as
follows:~Section~\ref{theory} describes  the
basic gates required for quantum computing with
qudits and the circuit decomposition of the qudit 
QFT.  Section~\ref{expt} describes the
experimental implementation of a hybrid qubit-qutrit QFT
on a four-qubit NMR quantum emulator.  Section~\ref{concl}
contains some concluding remarks.
\section{QFT Decompositions} 
\label{theory}
The state of a system of $N$ hybrid qudits 
each of a different dimension $d_p$
($p=1 \cdots N$)
can be written in terms of 
an orthonormal basis of product states
\begin{eqnarray}
\vert x \rangle &=& \vert x_{0} \rangle 
\otimes \vert x_{1} \rangle \otimes  \cdots  
\otimes \vert x_{N-1}\rangle  \nonumber \\
&\equiv& \vert x_{0} \cdots
x_{j} \cdots  x_{k}
\cdots x_{N-1}\rangle 
\end{eqnarray}
where $ x_{j} \in \{0,1,\dots d_{p}\}$.
For $d_p=d=2$, this reduces to an 
$N$ qubit state,
with $x =  \sum_{j=0}^{N-1} x_j 2^{j}$, 
and $ x_j \in \{0,1\}$
being binary integers. However, $x$ does not retain this
simple form for hybrid qudits with $d_p > 2$.
\subsection{Qudit Gates}
The Fourier gate ($F_p$) is a single-qudit gate that creates a
superposition of all basis states of the qudit, with equal
amplitudes, with its action on the
$p^{th}$ qudit in an $N$-qudit system given
by~\cite{daboul-jpa-03,muthu-jmo-02}: 
\begin{eqnarray}
F_{p} \vert x_{j} \rangle &=& \frac{1}{\sqrt{d}}\sum_{y_{k}=0}^{d-1}\exp 
\left(\frac{2\pi\iota x_{j}y_{k}}{d}\right) 
\vert y_{k} \rangle \nonumber \\
&& x_{j}, y_{k} \in [0,1,2, \dots (d-1)] 
\label{fourier}
\end{eqnarray}
For $d=2$ (qubit) and $d=3$ (qutrit) the Fourier gate  
reduces to the
Hadamard gate ($H$) and the Chrestenson gate 
($C$) respectively:
\begin{eqnarray}
H &=& \frac{1}{\sqrt{2}}\left(
\begin{array}{cc}
 1 & 1 \\
 1 & -1 \\
\end{array}
\right) 
\nonumber \\
C &=& \frac{1}{\sqrt{3}}\left(
 \begin{array}{ccc}
 1 & 1 & 1 \\
 1 & e^{\frac{2 \pi \iota}{3}} & e^{\frac{4 \pi
\iota}{3}} \\
 1 & e^{\frac{4 \pi \iota}{3}} & e^{\frac{2 \pi
\iota}{3}}
\end{array}
\right)
\end{eqnarray}
For a system of $N$ ``hybrid'' qudits, each of different
dimensions $d_{1}, d_{2}, \cdots d_{N}$, the
action of a two-qudit hybrid controlled-rotation gate
$R^{H}_{j,k}$ (with $j$ being the control qudit and $k$ the
target qudit) is given by~\cite{daboul-jpa-03,gottesman}:
\begin{eqnarray}
&&R^{H}_{j,k} \vert x_{0} \cdots x_{j}\cdots x_{k}
\cdots x_{N-1} \rangle = 
\nonumber \\
&&\quad\quad \exp \left( \frac{-2\pi\iota x_{j}x_{k}}
{\prod_{p=j}^{k}d_p}\right)
\vert x_{0} \cdots x_{j}\cdots x_{k} \cdots x_{N-1}
\rangle  \nonumber \\
\label{crotH}
\end{eqnarray}
where $d_{p}$ is the dimension of the $p^{th}$
qudit; $k > j$ has been assumed here, in order
to explicitly define the action of the gate
on the $N$-qudit basis state, however
interchanging $j$ and $k$ does not alter
the gate operation.
For $d_p=d=2$, the hybrid two-qudit controlled-rotation
gate reduces to the standard two-qubit controlled-rotation
gate $R_{j,k}$.
\subsection{Hybrid Qudit QFT}
The action of the QFT on the basis vectors of 
a hybrid qudit
system is given by~\cite{daboul-jpa-03,muthu-jmo-02}: 
\begin{equation}
{\rm QFT}
\ket{x}=\frac{1}{\sqrt{D}}\sum_{k=0}^{N-1}\exp
\left(\frac{2\pi
\iota xy}{D}\right)\ket{y}
\end{equation}
where the states $\vert y \rangle$ have the
same form as $\vert x \rangle$.

The circuit for implementing the QFT on $N$ hybrid qudits can be
decomposed as a set of single-qudit Fourier gates
interspersed with two-qudit controlled-rotation gates:\\
$F_{1} R^{H}_{1,2}
R^{H}_{1,3} \cdots R^{H}_{1,N} F_{2} R^{H}_{2,3} \cdots
F_{N-1} R^{H}_{N-1,N} F_{N}$. 
For $N$ qudits each of a different dimension $d_{j}$, the
total dimension of the Hilbert space is $D =
\prod_{j=1}^{N}d_{j}$, and it has to
be kept in mind while implementing
hybrid controlled-rotation gates that  
the dimensions of the control
and target qudits may be different.
After the implementation of the
QFT using the above decomposition, 
the bit values of the resultant state appear in 
the reverse
order. A sequence reversal can be
achieved either by applying a series of 
multi-valued permutation gates~\cite{muthu-pra-00}
or by reading off the
result in the reverse order.
\section{NMR Implementation}
\label{expt}
\subsection{The hybrid QQT system}
\label{emulate}
\begin{figure}[h]
\centering
\includegraphics[scale=1.0]{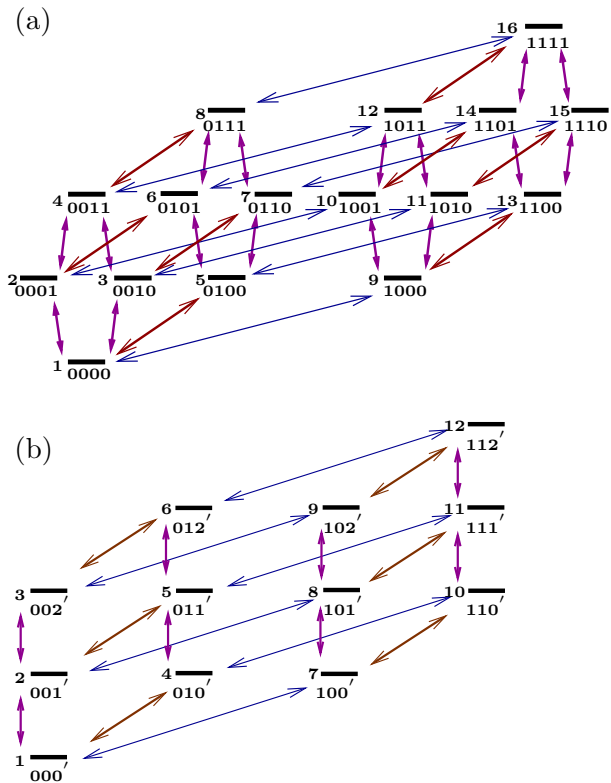}
\caption{
Energy level diagrams: (a) For a  
QQQQ system showing
16 energy levels corresponding to four coupled spin-$1/2$
particles. Spin states are given below each energy level,
$0$ and $1$ correspond to spin states $+\frac{1}{2}$ and
$-\frac{1}{2}$ respectively. There are 32 transitions
originating from single spin flips.  Transitions belonging
to spins 1 and 2 are shown in blue and brown respectively
while transitions belonging to spins 3 and 4 are shown in
purple.  (b) For a hybrid QQT system showing 12
energy levels and 20 transitions with energy levels labeled
in the computational basis; $0^{'}$, $1^{'}$ and $2^{'}$
correspond to the qutrit spin states $+1$, $0$ and $-1$
respectively. Each qubit undergoes six single spin flips
shown in blue (spin 1) and brown (spin 2) respectively.
Qutrit transitions are shown in purple.
\label{molecule}} 
\end{figure}
We consider a QQT system with a Hamiltonian 
governed by~\cite{slichter,levitt}
\begin{eqnarray}
\!\!\!\!\!\!\!\!\!\!\
H^{\rm QQT}=\omega_1 I_{1z}+\omega_2 I_{2z}+\Omega_3 I_{3z}+
D_Q(3I_{3z}^{2}-I^{2}) 
\nonumber \\
+J_{12}I_{1z}I_{2z} 
+ J_{23}^{\prime}I_{2z}I_{3z}+J_{13}^{\prime}I_{1z}I_{3z} 
\label{hamil}
\end{eqnarray}
where 1, 2 and 3 label the qubits and the
qutrit respectively, 
$I_{iz}$ is the z-component of the magnetization
vector of the $i^{th}$ qubit(qutrit), 
$I^{2}=\sum_{i} I_{3i}^{2}, \, i=x,y,z$
is the total magnetization of the qutrit, and $\omega_{1},
\omega_{2}, \Omega_{3}$
denote their Larmor frequencies i.e.
the free precession
of the qubits(qutrit) in a static magnetic field.
$D_{Q}$ represents the effective quadrupolar splitting of
the qutrit levels and the corresponding term in the
Hamiltonian accounts for the 
static first-order quadrupolar interaction.
$J_{ij}$s denote the strength of the scalar coupling interactions 
between $i^{th}$ and $j^{th}$ qubits, and
$J_{ij}^{\prime}$ represents an 
interaction term involving the
qutrit.

The QQT Hamiltonian in Eqn.~(\ref{hamil})
can be emulated by four qubits such
that two of the qubits mimic a three-level system. 
We use spins 3 and 4 of  our
four-qubit system to mimic a qutrit, such that 
the qutrit subspace containing levels 1,2 is emulated by the
third qubit and the qutrit subspace with levels 2,3
is emulated by the fourth qubit, with the 
corresponding chemical
shifts: $\omega_3=
\Omega_3+D_{Q}$ and $\omega_4=\Omega_3-D_{Q}$.

The Hamiltonian of a four-qubit QQQQ system 
(assuming a specific coupling 
pattern $J_{13}=J_{14}=J_{13}^{'}$,
$J_{23}=J_{24}=J_{23}^{'}$ and $J_{34}=0$)
is given by
\begin{eqnarray}
\!\!\!\!\!\!\!\!\!\!\
\!\!\!\!\!\!\!\
H^{\rm QQQQ}=\omega_1 I_{1z}+\omega_2 I_{2z}+\omega_3 I_{3z}+
\omega_4 I_{4z}+J_{12}I_{1z}I_{2z} \nonumber \\
+ J_{23}^{'}I_{2z}(I_{3z} +I_{4z}) 
+J_{13}^{'}I_{1z}(I_{3z}+I_{4z})
\end{eqnarray}
where $I_{iz}$ represents the z-component of the
magnetization vector of the $i^{th}$ spin.
The NMR spectrum and
multiplet pattern of $H^{\rm QQQQ}$ with the assumed $J$ coupling
values  resembles the first order NMR spectrum of a QQT
system (wherein each qubit multiplet contains six
transitions and the qutrit multiplet contains eight
transitions).  
Figure~1 shows the energy level diagram of the four-qubit
QQT emulator (energy levels numbered from 1-16) and an
actual QQT system (energy levels numbered from 1-12).

We use a 5-Fluorotryptophan molecule to emulate the QQT
system, with one $^{19}$F and three $^{1}$H nuclei
representing the four qubits. The molecular structure and
NMR parameters are depicted in Figure~2(a) and the thermal
equilibrium 1D NMR spectrum is shown in Figure~2(b).  The
T$_1$ and T$_2$ relaxation times for all four qubits range
between 1.17 - 3.96 s and 0.82 - 2.28 s respectively.  The
line intensities in the
coupling pattern of 5-Fluorotryptophan differs
slightly from that of the ideal QQT emulator, however hybrid qudit
gates implementation can be emulated on this system by
tailoring the $J-$evolution intervals to obtain the desired
angles of rotation.  The multiplet pattern and single
quantum NMR transitions  of the QQT emulator are shown in
Figure~3; labels below each spectral line refer to the
corresponding transitions as per Figure~1(a).  The mapping
between the transitions (labeled Q1, Q2, T) of the QQT
system and the four-qubit QQT emulator (labeled q1, q2, q3,
q4) is given in Table~1.  For an ideal QQT emulator, the
lines of the $i^{th}$ qubit occurring at $\omega_i \pm
J_{i3} \mp J_{i4}$ overlap with each other, while in the
5-Fluorotryptophan molecule, $J_{i3} \neq J_{i4}$($i \in
[1,2]$) and the lines occurring at $\omega_i \pm J_{i3} \mp
J_{i4}$ are separated by $2\mid J_{i3}-J_{i4}\mid$.  However
this molecule can emulate a QQT system whenever the
intensities of the transitions of the $i^{th}$ spin
occurring at $\omega_i \pm J_{i3} \mp J_{i4}$ are equal.
\begin{table}[h]
\label{table1}
\caption{Mapping between the transition frequencies
of a QQT system (labeled Q1, Q2 and T) 
and a four-qubit QQQQ system (labeled q1, q2, q3,
q4) which acts as QQT emulator.}
\begin{center}
\begin{tabular}{||l|l||l|l||l|l||}
\hline
\hline
Q1 & ~~q1 & Q2 & ~~q2 & ~T & q3, q4 \\
\hline 
\hline
1-7  & 1-9        & 1-4  &  1-5        & 1-2  &
1-3, 2-4 \\
                                              &&&&& \\
\hline
2-8  & 2-10 \&  & 2-5  &  2-6 \&    & 2-3  &  1-2, 3-4 \\
                                           &3-11 &&3-7&&\\
\hline
3-9  & 4-12       & 3-6  &  4-8        & 4-5  &
5-7, 6-8 \\
                                            &&&&& \\
\hline
4-10 & 5-13       & 7-10 &  9-13       & 5-6  &
5-6,7-8  \\
                                            &&&&&  \\
\hline
5-11 & 6-14 \& & 8-11 & 10-14 \& & 7-8  &  9-11, 10-12\\
                                           &7-15&&11-15&&  \\
\hline
6-12 & 8-16       & 9-12 & 12-16       & 8-9  &
9-10, 11-12 \\
                                                  &&&&&  \\
\hline
  -- & --         & --   &  --         &10-11 &
13-15, 14-16\\
                                                 &&&&& \\
\hline
  -- & --         & --   &  --         &11-12 &
13-14, 15-16 \\
                                                 &&&&&  \\
\hline
\hline
\end{tabular} 
\end{center}
\end{table}
\begin{figure}[h]
\centering
\includegraphics[scale=1.0]{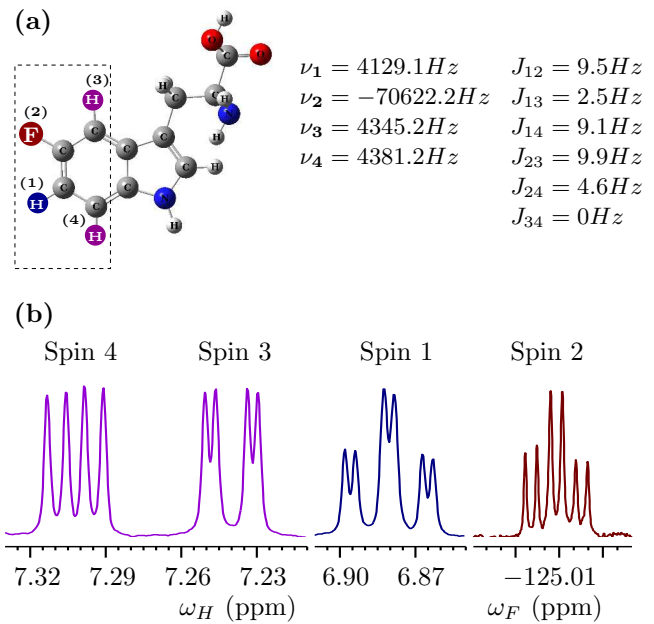}
\caption{(a) Molecular structure of
5-Fluorotryptophan with qubits of interest shown
in the rectangular block and numbered from 1-4.
Three $^{1}$H nuclei and one $^{19}$F nucleus form
the four-qubit system. The
chemical shift values (in ppm) and indirect
coupling constants (in Hz) are tabulated
alongside. (b) NMR spectrum of
the thermal state of 5-Fluorotryptophan. The
${}^{1}$H and ${}^{19}$F spectra are shown on the same scale. Colors of
the spectral lines correspond to the 
spin color given in the molecular structure.
The spins resonate at the
frequencies $\omega_1$, $\omega_2$
and $\omega_3$, with $J_{ij}$ corresponding to
the scalar coupling
constant between spins $i,j$.
\label{5ftrp}}
\end{figure}
\subsection{Experimental implementation of the QFT}
The NMR pulse sequence for the experimental implementation
of a hybrid qudit QFT on the four-qubit QQT NMR emulator is
shown in Figure~4.  A hybrid controlled-rotation gate
$R^{H}_{j,k}$ (where `$j$' is the control-qubit and `$k$' is
the target qutrit) can be realized by a z-rotation of the
qutrit state by an angle $\frac{2\pi x_j
x_k}{\prod_{m=j}^{k}d_{m}}$ ($j \in [0,1], \, k \in [0,1,2]$).
For the two qubits mimicking the qutrit, this
controlled-rotation can be obtained by rotating both the
qubits by the same angle with respect to the control qubit.
Chemical shift offsets are chosen such  that spins 1 and 2
always remain on resonance.

The action of the QFT in extracting
state periodicity can be equally well demonstrated on an
initial thermal equilibrium state as on a pseudopure
state and we hence implement the QFT on a thermal
equilibrium state~\cite{weinstein-prl-01}.
Beginning with the thermal equilibrium density operator, the
first $\left(\frac{\pi}{2}\right)^{1}_{\mathbf{n_1}}$ pulse implements
a Hadamard gate on the first qubit. The axis of rotation
$\mathbf{n_{1}}=\cos{\theta_1} \hat{x} +
\sin{\theta_1}\hat{y}$ (aligned at
an angle of $\theta_1=\frac{\pi}{2}+\frac{5\pi}{6}$ with the
x-axis) was chosen to obtain  a $z-$rotation of the first
qubit by an angle $\frac{5\pi}{6}$, 
thereby circumventing the
high time cost involved in implementing $z-$rotations.
At this point in the pulse sequence, the
qubits 2, 3 and 4 do not have any coherence, and thus single
spin $z-$rotations are redundant in their case.  The first
evolution period incorporates the controlled-rotation 
gates $R_{1,2}$ and $R^{H}_{1,3}$ for a time
$\tau_{ij}=\frac{\theta_{ij}}{\pi J_{ij}}$ ($\theta_{ij}$ is
the desired rotation angle).  The evolutions under
the scalar coupling $J_{12}$ by an angle $\frac{\pi}{2}$, 
under the scalar coupling $J_{13}$
by an angle $\frac{\pi}{6}$, and 
under the scalar coupling $J_{14}$ by an angle
$\frac{\pi}{6}$ are achieved simultaneously during the first
evolution period.  Refocusing pulses are introduced at
appropriate points to obtain the desired rotation angles.
The next module of the hybrid QFT begins with a
$\left(\frac{\pi}{2}\right)_{\mathbf{n_2}}^{2}$ pulse that implements a Hadamard
gate along with producing an effective
$z-$rotation of $\frac{2\pi}{3}$ on the second qubit (the
axis $\mathbf{n_{2}}$ is chosen to make an angle of
$\theta_2=\frac{\pi}{2}+\frac{2\pi}{3}$ with the x-axis).
This is followed by an evolution under the
couplings $J_{23}$
and $J_{24}$ during the second evolution interval. 
The
effect of evolution on the first spin is refocused by a
pulse in the middle of the evolution period.  The
final Hadamard gate on the third and fourth qubits is performed
by a $\left(\frac{\pi}{2}\right)_y^{3,4}$ pulse (which effectively
mimics a single-qutrit Fourier gate).
All pulses on qubit 1 are Gaussian
pulses of 13 ms duration and the simultaneous excitation of
qubits 3 and 4 is achieved using a 7 ms shaped pulse
({\em Seduce.100}).  All the pulses on qubit 2 are hard pulses.
Since qubits 3 and 4 are only 36 Hz apart it is difficult to
individually address them in an experiment, and hence the
controlled-rotation gates on these two qubits have been
implemented without applying individual refocusing pulses on
them.
\begin{figure}[h]
\centering
\includegraphics[scale=0.8]{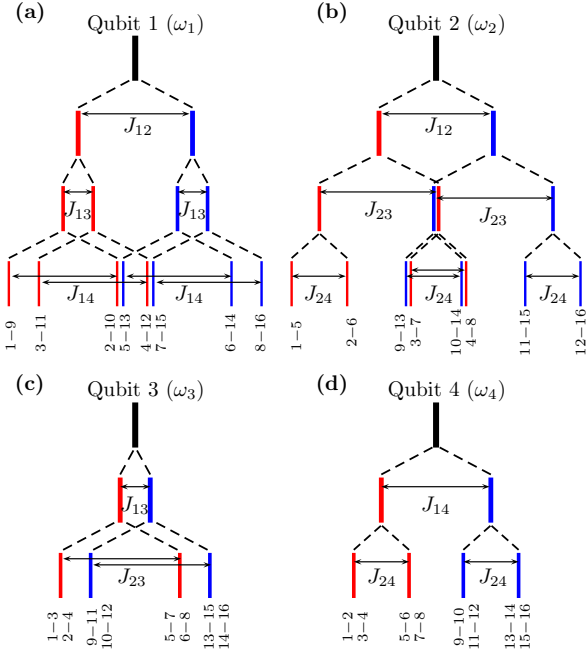}
\caption{Multiplet pattern: (a) Of qubit 1, (b) Of qubit 2, (c)
Of qubit 3 and (d) Of qubit 4 of the four-qubit NMR emulator.
Each multiplet is centered around its chemical shift value
(black lines representing $\omega_1, \omega_2, \omega_3$ and
$\omega_4$ respectively), which is further split by scalar
coupling interaction.  The ratio of coupling constants
$J_{12}:J_{13}:J_{14}$, $J_{12}:J_{23}:J_{24}$,
$J_{13}:J_{23}$ and $J_{14}:J_{24}$ is the same as in the
5-Fluorotryptophan molecule.  The line thickness is
proportional to the spectral line intensity.  Labels below each line refer
to the corresponding transition as marked in Figure~1(a).
\label{figtree}}
\end{figure}
\begin{figure}[h]
\centering
\includegraphics[scale=1.0]{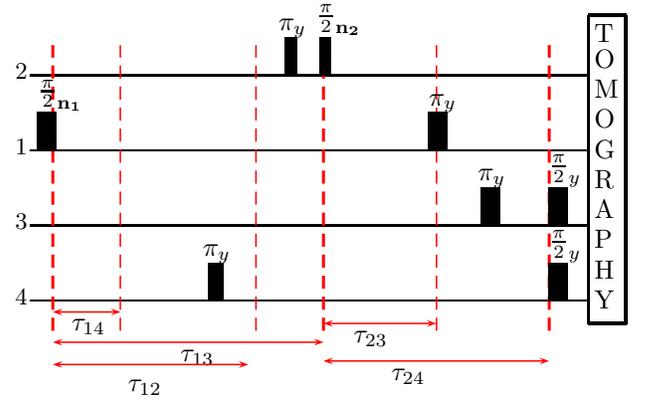}
\caption{
Pulse sequence for a hybrid QFT on a qubit-qubit-qutrit
(QQT) system implemented on a four-qubit emulator.  Flip
angles and axes of rotation are displayed over each pulse.
Rotation axes $\mathbf{n_1}$ and $\mathbf{n_2}$ are aligned at angles
$\theta_1=\frac{\pi}{2}+\frac{5\pi}{6}$ and
$\theta_2=\frac{\pi}{2}+\frac{2\pi}{3}$ with respect to the
x-axis. Thick black rectangles  represent selective pulses
while thin black rectangles represent non-selective pulses;
$\tau_{ij}$s represent time intervals of evolution under the
scalar couplings $J_{ij}$.
\label{pulse}}
\end{figure}
\subsection{Tomographic reconstruction}
Partial tomography of the QQQQ system was performed in order to
reconstruct the desired QQT density matrix.  
First, a set of operators were
designed to reconstruct  the  full QQT density operator in $d
\times d$ dimensional Liouville space (in this case $d=2
\otimes 2 \otimes 3$).  This set of operators was
then extended to the analogous set of tomography operations in a
$16 \times 16$ dimensional operator space of the four-qubit
NMR emulator.

The complete characterization of a $12 \times 12$
dimensional QQT density operator requires the determination
of 143 variables.  The 11 diagonal elements (populations)
are obtained by applying an appropriate z-gradient to kill
off-diagonal elements, followed by spin-selective rotations
to project the diagonal elements onto experimentally
measurable parts of the density matrix.  The remaining 132
elements are obtained by a set of 19 operators:
III$^{\prime}$, YII$^{\prime}$, XII$^{\prime}$,
IYI$^{\prime}$, IXI$^{\prime}$, IIY$^{\prime}$,
IYY$^{\prime}$, IXY$^{\prime}$, YIY$^{\prime}$,
XIX$^{\prime}$, XXI$^{\prime}$, YYI$^{\prime}$,
XYI$^{\prime}$, YXI$^{\prime}$, XYY$^{\prime}$,
XXY$^{\prime}$, YXX$^{\prime}$, YYX$^{\prime}$ and
II$\Lambda^{\prime}$-{\bf Grad}$_z$-II$\Upsilon^{\prime}$,
where I is the identity operator,   X(Y) refers to a single
spin operator and primed operators correspond to qutrit
operations.  These operators can be implemented by applying
the corresponding spin-selective $\frac{\pi}{2}$ pulses (or
a no-operation for the identity operator).  The last
tomography experiment consists of a $z$ gradient pulse
sandwiched between two $y$ pulses 
represented by $\Lambda$ and $\Upsilon$ 
applied on the qutrit,
of flip angles
$(-\frac{\pi}{2})$ and $(\frac{\pi}{4})$ respectively.

A corresponding set of 19 operations can be used to
determine the off-diagonal elements of the four-qubit QQQQ
system: IIII, YIII, XIII, IYII, IXII, IIYY, IYYY, IXYY,
YIYY, XIXX, XXII, YYII, XYII, YXII, XYYY, XXYY, YXXX, YYXX
and II$\Lambda \Lambda$-{\bf Grad}$_z$-II$\Upsilon
\Upsilon$.  The last tomography experiment on the QQQQ
system consists of a $z$ gradient pulse sandwiched between
two $y$ pulses  represented by $\Lambda$ and $\Upsilon$,
applied simultaneously on qubits 3 and 4 of flip angles
$(-\frac{\pi}{2})$ and $(\frac{\pi}{4})$ respectively. 

The theoretically expected and experimentally
obtained  tomographs of the real and imaginary parts of the
density matrices obtained after applying a hybrid QFT on the
four-qubit QQT NMR emulator are given in Figure~5.  
On visual inspection, one can see that the general pattern
of the weights in the density matrices 
is quite similar, indicating the success of the
QFT implementation. Quantitatively, the fidelity of the
reconstructed state was computed using the 
measure~\cite{weinstein-prl-01}:
\begin{equation}
F =
\frac{Tr(\rho_{\rm theory}^{\dag}\rho_{\rm expt})}
{\sqrt(Tr(\rho_{\rm theory}^{\dag}\rho_{\rm theory}))
\sqrt(Tr(\rho_{\rm expt}^{\dag}\rho_{\rm expt}))}
\label{fidelity}
\end{equation}
where $\rho_{\rm theory}$ and $\rho_{\rm expt}$ denote the
theoretical and experimental density matrices
respectively.
The fidelity of the experimentally reconstructed density
matrix was computed to be 0.83.  The loss in fidelity of the
experimentally reconstructed state can be attributed to
several experimental factors, namely, rf pulse
imperfections, imperfect refocusing of chemical shift
evolution of unwanted coherences during the evolution
intervals, rf field inhomogeneity, and deleterious effects
due to spin relaxation.  Our experiments point the way to an
interesting direction of research in quantum computing with
NMR namely, that of using complex coupled qubit topologies to
emulate hybrid and higher-dimensional quantum systems.
\begin{figure}[h]
\centering
\includegraphics[scale=1.0]{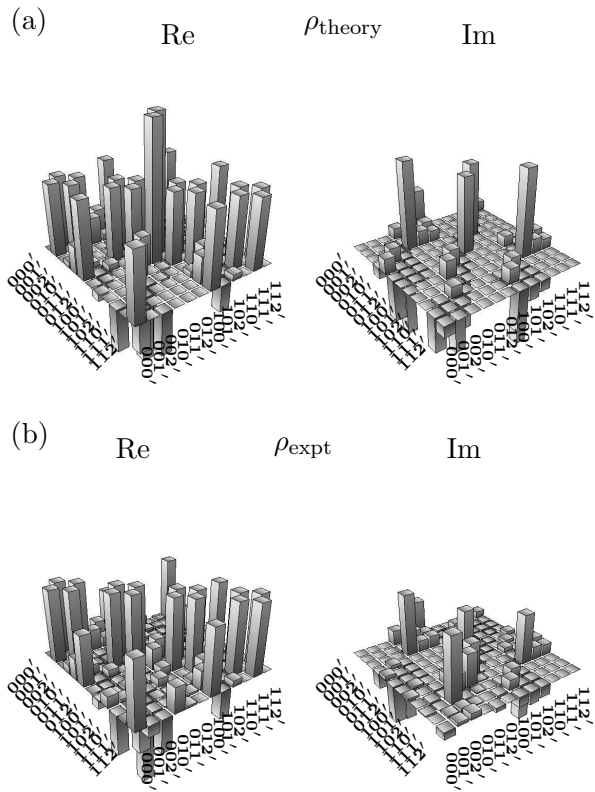}
\caption{Tomographs of the real and imaginary
parts of the (a) theoretically expected ($\rho_{\rm theory}$) and 
(b) experimentally obtained ($\rho_{\rm expt}$)
final density operator  
obtained after applying a hybrid QFT on the thermal 
equilibrium state of the four-qubit QQQQ system
used as a QQT emulator. The rows and columns encode the
computational basis in binary order, from
$\vert 0 0 0^{\prime} \rangle$ to $\vert 1 1 2^{\prime}
\rangle$.
\label{tomo-exp}
}
\end{figure}
\section{Concluding remarks} 
\label{concl}
It has been proposed that qudit-based quantum computers are
able to better optimize Hilbert-space dimensionality and are
hence expected to be more powerful than the standard models
of qubit quantum computers.  One of the key quantum circuits
in several quantum algorithms is the QFT and it is hence
important to look for decompositions of the QFT specifically
designed for qudits and hybrid qubit-qudit systems.  We used
four NMR qubits to emulate a hybrid system of two qubits
coupled to a qutrit and implemented the QFT on this hybrid
quantum computer.  It is expected that these experiments
will pave the way for the implementation of full-fledged
qudit-based quantum computing proposals.

\vspace*{1cm}
\noindent{\bf Acknowledgments}
All experiments were performed on a Bruker 600 MHz FT-NMR
spectrometer in the NMR Research Facility at IISER Mohali.
SD acknowledges UGC India for financial support.

\vspace*{1cm}

\begin{thebibliography}{10}
\expandafter\ifx\csname url\endcsname\relax
  \def\url#1{\texttt{#1}}\fi
\expandafter\ifx\csname urlprefix\endcsname\relax\def\urlprefix{URL }\fi
\expandafter\ifx\csname href\endcsname\relax
  \def\href#1#2{#2} \def\path#1{#1}\fi

\bibitem{greentree-prl-04}
A.~D. Greentree, S.~G. Schirmer, F.~Green, L.~C.~L. Hollenberg, A.~R. Hamilton,
  R.~G. Clark, Maximizing the Hilbert space for a finite number of
  distinguishable quantum states, Phys.Rev.Lett. 92 (2004) 097901.

\bibitem{lanyon-naturephy-09}
B.~P. Lanyon, M.~H. Barbieri, M.~P. Almeida, T.~Jennewein, T.~C. Ralph, K.~J.
  Resch, G.~J. Pryde, J.~L. O'Brien, A.~Gilchrist, A.~G. White, Simplifying
  quantum logic using higher-dimensional Hilbert spaces,
Nature Phys. 5 (2009)
  134--140.

\bibitem{brennen-pra-05}
G.~K. Brennen, D.~P. O'Leary, S.~S. Bullock, Criteria for exact qudit
  universality, Phys. Rev. A 71 (2005) 052318.

\bibitem{klimov-pra-03}
A.~B. Klimov, R.~Guzm\'an, J.~C. Retamal, C.~Saavedra, Qutrit quantum computer
  with trapped ions, Phys.Rev.A 67 (2003) 062313.

\bibitem{lin-pra-09}
Q.~Lin, B.~He, Bi-directional mapping between polarization and spatially
  encoded photonic qutrits, Phys. Rev. A 80 (2009) 062312.

\bibitem{muthu-pra-00}
A.~Muthukrishnan, C.~R. Stroud, Multivalued logic gates for quantum
  computation, Phys. Rev. A 62 (2000) 052309.

\bibitem{brennen-qic-2006}
G.~K. Brennen, S.~S. Bullock, D.~P. O'Leary, Efficient circuits for
  exact-universal computations with qudits, Quant. Inf.
Comp.
  6 (2006) 436.

\bibitem{bullock-prl-2005}
S.~S. Bullock, D.~P. O'Leary, G.~K. Brennen, Asymptotically optimal quantum
  circuits for $d$-level systems, Phys. Rev. Lett. 94 (2005) 230502.

\bibitem{wilmott-ijqi-12}
C.~M. Wilmott, P.~R. Wild, On a generalized quantum swap gate, Intl. J. Qtm.
  Inf. 10 (2012) 1250034.

\bibitem{di-pra-13}
Y.-M. Di, H.-R. Wei, Synthesis of multivalued quantum logic circuits by
  elementary gates, Phys.Rev.A 87 (2013) 012325.

\bibitem{garcia-qip-13}
J.~C. Garcia-Escartin, P.~Chamorro-Posada, A swap gate for qudits, Qtm. Inf.
  Proc. 12 (2013) 3625.

\bibitem{khan-comp-06}
F.~S. Khan, M.~Perkowski, Synthesis of multi-qudit hybrid and d-valued quantum
  logic circuits by decomposition, Theor. Comp. Sci. 367 (2006) 336.

\bibitem{li-pra-13}
W.-D. Li, Y.-J. Gu, K.~Liu, Y.-H. Lee, Y.-Z. Zhang, Efficient universal quantum
  computation with auxiliary Hilbert space, Phys.Rev.A 88 (2013) 034303.

\bibitem{rousseaux-pra-13}
B.~Rousseaux, S.~Gu\'erin, N.~V. Vitanov, Arbitrary qudit gates by adiabatic
  passage, Phys.Rev.A 87 (2013) 032328.

\bibitem{daboul-jpa-03}
J.~Daboul, X.~Wang, B.~C. Sanders, Quantum gates on hybrid qudits, J.Phys.A 36
  (2003) 2525.

\bibitem{gottesman}
D.~Gottesman, Fault-tolerant quantum computation with higher-dimensional
  systems, Chaos, Solitons and Fractals 10 (1999) 1749.

\bibitem{peres-prl-00}
H.~Bechmann-Pasquinucci, A.~Peres, Quantum cryptography with 3-state systems,
  Phys.Rev.Lett. 85 (2000) 3313.

\bibitem{groblacher-njp-06}
S.~Groeblacher, T.~Jennewein, A.~Vaziri, G.~Weihs, A.~Zeilinger, Experimental
  quantum cryptography with qutrits, New J. Phys. 8 (2006) 75.

\bibitem{leary-pra-06}
D.~P. O'Leary, G.~K. Brennen, S.~S. Bullock, Parallelism for quantum
  computation with qudits, Phys. Rev. A 74 (2006) 032334.

\bibitem{nagali-prl-10}
E.~Nagali, D.~Giovannini, L.~Marrucci, S.~Slussarenko, E.~Santamato,
  F.~Sciarrino, Experimental optimal cloning of four-dimensional quantum states
  of photons, Phys.Rev.Lett. 105 (2010) 073602.

\bibitem{das-ijqi-2003}
R.~Das, A.~Mitra, V.~S. Kumar, A.~Kumar, Quantum information
processing by NMR:
  preparation of pseudopure states and implementation of unitary operations in
  a single-qutrit system., Intl. J. Qtm. Inf. 1~(3) (2003) 387--394.

\bibitem{pinto-pra}
D.~O. Soares-Pinto, L.~C. Celeri, R.~Auccaise, F.~F. Fanchini, E.~R. deAzevedo,
  J.~Maziero, T.~J. Bonagamba, R.~M. Serra, Nonclassical
correlations in NMR
  quadrupolar systems, Phys.Rev.A 81 (2010) 062118.

\bibitem{dogra-pla-14}
S.~Dogra, Arvind, K.~Dorai, Determining the parity of a permutation using an
  experimental NMR qutrit, Phys. Lett. A 378 (2014) 3452.

\bibitem{pinto-qph}
I.~A. Silva, B.~Cakmak, G.~Karpat, E.~L.~G. Vidoto, D.~O. Soares-Pinto, E.~R.
  deAzevedo, F.~F. Fanchini, Z.~Gedik, Computational speed-up in a single qudit
  NMR quantum information processor, E-print,\href
  {http://arxiv.org/abs/1406.3579} {\path{arXiv:1406.3579}}.

\bibitem{khitrin-jcp}
A.~K. Khitrin, B.~M. Fung, Nuclear magnetic resonance quantum logic gates using
  quadrupolar nuclei, J. Chem. Phys. 112 (2000) 6963.

\bibitem{neeraj-jcp}
N.~Sinha, T.~S. Mahesh, K.~V. Ramanathan, A.~Kumar, Toward quantum information
  processing by nuclear magnetic resonance: Pseudopure states and logical
  operations using selective pulses on an oriented spin 3/2 nucleus, J. Chem.
  Phys. 114 (2001) 4415.

\bibitem{shor-sjc-1997}
P.~W. Shor, Polynomial time algorithms for prime factorization and discrete
  logarithms on a quantum computer, SIAM J. Comput. 26 (1997) 1484.

\bibitem{cleve-proc}
R.~Cleve, A.~Ekert, C.~Macchiavello, M.~Mosca, Quantum algorithms revisited,
  Proc. Roy. Soc. London A. 454 (1998) 339.

\bibitem{sim-sjc-1997}
D.~R. Simon, On the power of quantum computation, SIAM J. Comput. 26 (1997)
  1474.

\bibitem{coppersmith-94}
D.~Coppersmith, An approximate Fourier transform useful in quantum factoring,
  IBM Research Report (1994) RC19642.

\bibitem{weinstein-prl-01}
Y.~S. Weinstein, M.~A. Pravia, E.~M. Fortunato, S.~Lloyd, D.~G. Cory,
  Implementation of the quantum Fourier transform, Phys.Rev.Lett. 86 (2001)
  1889.

\bibitem{dorai-ijqi-05}
K.~Dorai, D.~Suter, Efficient implementations of the quantum
Fourier transform:
  An experimental perspective, Intl. J. Qtm. Inf. 3 (2005) 413.

\bibitem{wang-jphysb-11}
H.-F. Wang, X.-X. Jiang, S.~Zhang, K.-H. Yeon, Efficient quantum circuit for
  implementing discrete quantum Fourier transform in solid-state qubits, J.
  Phys. B. 44 (2011) 115502.

\bibitem{obada-josab-13}
A.-S. Obada, H.~A. Hessian, A.-B. Mohamed, A.~H. Homid, Implementing discrete
  quantum Fourier transform via superconducting qubits coupled to a
  superconducting cavity, J. Opt. Soc. Am. B. 30 (2013) 1178.

\bibitem{dong-josab-13}
L.~Dong, X.-M. Xiu, H.-Z. Shen, Y.-J. Gao, X.~X. Yi, Quantum
Fourier transform
  of polarization photons mediated by weak cross-Kerr nonlinearity, J. Opt.
  Soc. Am. B. 30 (2013) 2765.

\bibitem{muthu-jmo-02}
A.~Muthukrishnan, C.~R. Stroud, Quantum fast Fourier transform using multilevel
  atoms, J. Mod. Opt. 49 (2002) 2115.

\bibitem{zobov-jetp-06}
V.~E. Zobov, A.~S. Ermilov, Pulse sequences for realizing
the quantum Fourier
  transform on multilevel systems, JETP Lett. 83 (2006) 467.

\bibitem{zilic-ieee-07}
Z.~Zilic, K.~Radecka, Scaling and better approximating
quantum Fourier
  transform by higher radices, IEEE Trans. Comp. 56~(2) (2007) 202--207.

\bibitem{ye-comm-11}
Y.~Cao, S.~G. Peng, C.~Zheng, G.~L. Long, Quantum Fourier transform and phase
  estimation in qudit system, Commun. Theor. Phys. 55 (2011) 790.

\bibitem{gopinath-pra-2006}
T.~Gopinath, A.~Kumar, Geometric quantum computation using fictitious spin- 1/2
  subspaces of strongly dipolar coupled nuclear spins, Phys.Rev.A 73 (2006)
  022326.

\bibitem{lee-apl-2006}
J.~S. Lee, A.~K. Khitrin, Projective measurement in nuclear magnetic resonance,
  App. Phys. Lett. 89 (2006) 074105.

\bibitem{khitrin-pra}
A.~Khitrin, H.~Sun, B.~M. Fung, Method of multifrequency excitation for
  creating pseudopure states for NMR quantum computing, Phys.Rev.A 63 (2001)
  020301R.

\bibitem{neeraj-pra}
K.~V. R.~M. Murali, N.~Sinha, T.~S. Mahesh, M.~H. Levitt, K.~V. Ramanathan,
  A.~Kumar, Quantum-information processing by nuclear magnetic resonance:
  Experimental implementation of half-adder and subtractor operations using an
  oriented spin-7/2 system, Phys.Rev.A 66 (2002) 022313.

\bibitem{pinto-ijqi}
A.~G. Araujo-Ferreira, C.~A. Brasil, D.~O. Soares-Pinto, E.~R. deAzevedo, T.~J.
  Bonagamba, J.~Teles, Quantum state tomography and quantum logical operations
  in a three qubits NMR quadrupolar system, Intl. J. Qtm. Inf. 10 (2012)
  1250016.

\bibitem{slichter}
C.~P. Slichter, Principles of magnetic resonance, Springer, Newyork, 1996.

\bibitem{levitt}
M.~H. Levitt, Spin dynamics : Basics of nuclear magnetic resonance, John Wiley
  and Sons, Chichester England, 2008.
\end{thebibliography}

\end{document}